\documentclass[12pt,preprint]{aastex}
\usepackage{subfigure}
\usepackage{natbib}
\usepackage{graphicx}
\shorttitle{first photometric study of LO Com}
\shortauthors{Zhang et al.}
\begin{document}

\title{The first photometric study of the short period shallow contact system LO Comae}

\author{Y. Zhang,\altaffilmark{1} Q. W. Han, \altaffilmark{2,3,4} and  J. Z. Liu\altaffilmark{1}}

\altaffiltext{1}{Xinjiang Astronomical Observatory,
Chinese Academy of Sciences, Urumqi 830011, China;
zhy@xao.ac.cn}
\altaffiltext{2}{Yunnan Observatory, Chinese Academy of Sciences, Kunming 650011, China}
\altaffiltext{3}{Key
Laboratory for the Structure and Evolution of Celestial Objects,
Chinese Academy of Sciences, Kunming 650011, China}
\altaffiltext{4}{University of Chinese Academy of Sciences, Beijing 100049, China}

\begin{abstract}\label{abstract}
%\begin{abstract}
In this paper, the first complete photometric light curves in the $B$, $V$, and $R$ passbands for an eclipsing binary LO Com are presented, and the photometric solution for the LO Com is derived by analyzing these light curves by using the Wilson and Devinney code. From the photometric solution, it is found that the LO Com is a W-type W UMa binary with a mass ratio of $q=m_{2}/m_{1}=2.478$ and a contact degree of $f = (3.2\pm0.25)\%$. By combining the two new minimum times with those published earlier in the literature, we have found that the $(O-C)$ curve shows a downward parabolic variation  corresponding to a long-term decrease in the orbital period with a rate of ${\rm d}P/{\rm d}t=-1.18\times10^{-7}$days yr$^{-1}$. This long-term decrease in its orbital period may be caused by mass transfer from the more massive component to the less massive one. 
\end{abstract}


\section{Introduction}
\label{sect:intro}
W Ursae Majoris (W UMa) binaries are eclipsing variables in which both components have filled their Roche lobes and shared a common envelope. As the most common observed variables, their eclipsing light curves have nearly equal depths. W UMa systems are important sources for testing the angular momentum evolution of binaries \citep{Rucinski2000}. They have a high spatial frequency of occurrence and can also be used as standard candle for distance determinations \citep{Rucinski1997AJ}. They play an important role in studying the Galaxy structure. More importantly, the two components in W UMa systems transfer the mass and energy between themselves. They are good targets for understanding the energy and mass transfer mechanism \citep{Li2004}. On the other hand, they also play an important role in investigating the stellar evolution as they are the possible progenitors for some important objects, e.g. blue stragglers \citep{Chen2008} and fast rotating stars \citep{Jiang2013}.

The majority of contact binary systems with short periods belong to W-subtype systems \citep{Webbink2003ASPC}, and in this system the more massive component is the cooler one. Most of the W-subtype contact systems have shallow contact characteristics. Thus, they are the excellent objects for testing the thermal relaxation oscillation (TRO) model \citep[][]{Lucy1976,Flannery1976,Robertson1977,Lucy1979,Li2004,Li2005}. In the TRO model, the W UMa binaries oscillate between contact and semi-detached states.
In the semi-detached phase, the primary continues to fill its Roche lobe and the mass transfer from the primary to the secondary, whereas no energy transfers between them. In the contact phase, energy transfers from the primary to the secondary, and the mass transfers from the secondary to the primary. For the case that a W UMa binary has low shallow contact and keeps expanding, the contact configuration can break, and the system will evolve into the semi-detached state. Therefore, the short shallow contact systems are important targets to test the TRO theory.

The orbital periods of W UMa binaries are between $0.22$ and $1.00$ days \citep{Gazeas2006MNRAS}. The short-period W UMa binary LO Com (GSC 1991-1390, 2MASS J12320490+2622477, NSVS 7622465; $\rm \alpha_{2000}= 12^{h}32^{m}04^{s}.91$, $\rm \delta_{2000}= +26^{\circ}22^{'}47^{''}.7$) is classified as an EW-type eclipsing binary and is poorly investigated. Even though \citet{Blattler2001IBVS} presented the light curves for this system, no complete photometric analysis has been published. In this paper, we show as the first time complete photometric or charge-coupled device (CCD) light curves in the $B$, $V$, and $R$ passbands for the LO Com in Section 2. Then, the photometric solution with the Wilson and Devinney (W-D) program and analysis of the orbital period are presented in Sections 3 and 4, respectively. Finally, the discussions based on the photometric solution and orbital period variation are presented in Section 5.

\section{Observations}
\label{sect:observe}
CCD observations of the LO Com were taken on 2015 March $21$ and $25$ at the Nanshan $1.0$ m  telescope of Xinjiang Astronomical Observatory. The telescope is equipped with a $4$K ${\times}$ $4$K camera CCD. During the observation, the Johnson-Cousins $BVR$ filters were used.
The images were reduced with the aperture photometry package of IRAF.\footnote{IRAF is distributed by the National Optical Astronomy Observatory, which is operated by the Association of Universities for Research in Astronomy, Inc., under the cooperative agreement with the National Science Foundation.} Figure 1 shows one of the observed CCD images containing the variable stars, the comparison star, and the check star. Table 1 presents the coordinates of the variable, comparison, and check stars. Our observation data of three passbands are listed in Table 6, in delta magnitudes, the variable star minus the comparison star.
\begin{figure}
\begin{center}
 \includegraphics[bb=85 480 370 800,height=17.cm,width=15.cm,clip,angle=0,scale=0.5,angle=0]{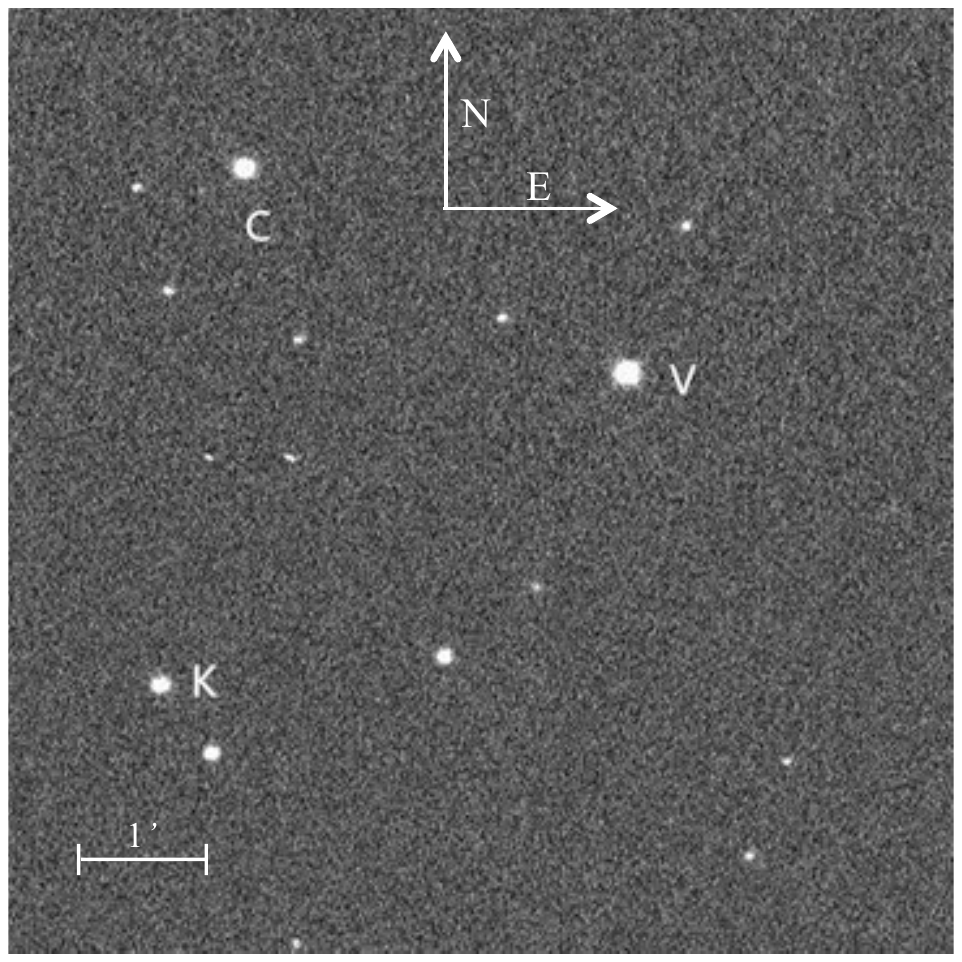}
 \caption{A CCD image of the LO Com V, C, K indicate the variable, comparison, and check stars, respectively. }
 \label{fig:photoz_com}
\end{center}
\end{figure}

\begin{deluxetable}{lccccc}
\label{table:1}
\tablecolumns{4}
\tablewidth{0pc}
\tablecaption{The coordinates of the variable, Comparison, and Check stars}
\tablehead{

\colhead{Objects} & \colhead{Name}   & \colhead{R.A.(2000)}    & \colhead{DEC.(2000)} & \colhead{$B$}  & \colhead{($B-R$)}}
\startdata

Variable   & LO Com                     & 12:32:04.91    & +26:22:47.7  &14.0 &1.5\\
Comparison & Arty 804                   & 12:31:52.80    & +26:24:23.7  &14.3 &1.4\\
Check      & 2MASS J12314932+2620407    & 12:31:49.32    & +26:20:40.7  &15.4 &1.2\\

\enddata
\end{deluxetable}

The complete light curves in the $BVR$ passbands are presented in the top panel of Figure 2. The orbital phases of these observations were obtained by
$Min. I=2457107.2343+0^{\rm d}.2863601\times E$,
where the $2457107.2343$ is one of the new times of light minima obtained by us and the period is adopted from \citet{Blattler2001IBVS}. The magnitude differences between the comparison and check stars in the $BVR$ passbands are shown in the bottom panel of Figure 2; the bottom panel shows the authenticity of the variations of the light curves for the target. The standard observation accuracies are about $0.021, 0.016$, and $0.011$ for the $B, V$, and $R$ bands, respectively. From Figure 2, it can be seen that the light curves are typically the EW-type, which are symmetric in all passbands and have no O'Connell effects (different heights of the two light maxima). Figure 2 also shows that there is $\sim0.13$ mag for three passbands between the depths of the two minima.  From the observations, two times of the  minimum light were derived for each passband by using the method of \citet{Kwee1956BAN} listed in Table 2.

\begin{deluxetable}{lccc}
\label{table:1}
\tablecolumns{4}
\tablewidth{0pc}
\tablecaption{New times of light minimum of the LO Com.}
\tablehead{

\colhead{HJD} & \colhead{Error}   & \colhead{Type}    & \colhead{Filter} }
\startdata
2457107.2343   & 0.0001  & I  & B \\
2457107.2344   & 0.0002  & I  & V \\
2457107.2342   & 0.0001  & I  & R \\
2457103.3682   & 0.0003  & II  & B \\
2457103.3682   & 0.0002  & II  & V \\
2457103.3684   & 0.0001  & II  & R \\
\enddata
\end{deluxetable}
\begin{figure}
\begin{center}
 \includegraphics[bb=40 40 780 600,height=18.cm,width=25.cm,clip,angle=0,scale=0.5,angle=0]{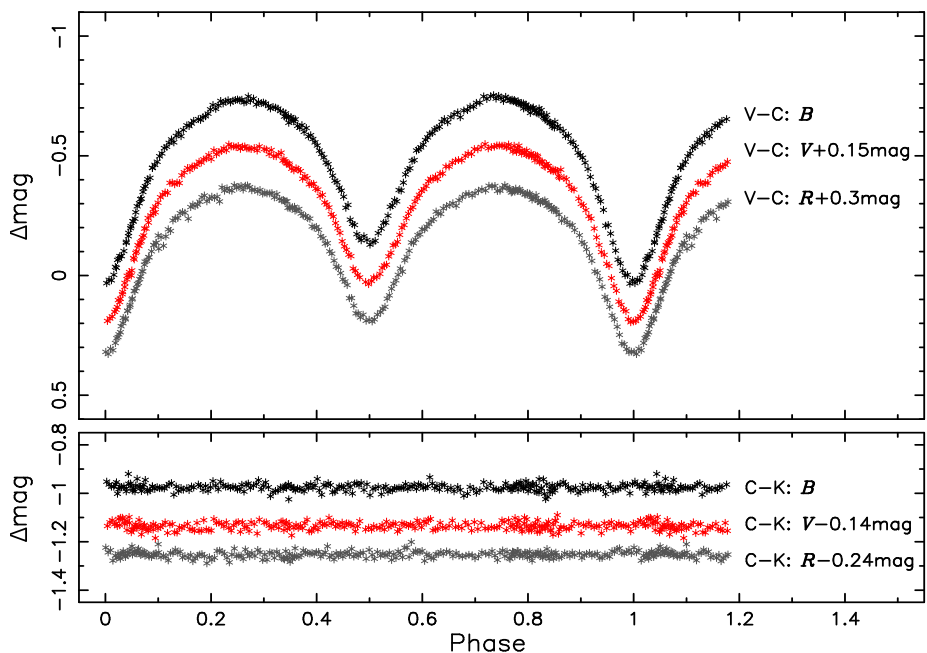}
 \caption{Top panel shows the CCD photometric light curves in the $BVR$ passbands for the LO Com. The black, red, and gray asterisks indicate for the $B, V$ and $R$ passbands, respectively. Note that the curve in the $V$ and $R$ passbands are shifted by $+0.15$ and $+0.30$ mag, respectively.  In the bottom panel, the magnitude differences between the comparison and check stars are shown. Here, shifts of $-0.14$ and $-0.24$ mag are also made for the $V$ and $R$ passbands, respectively.}
 \label{fig:photoz_com}
\end{center}
\end{figure}

\begin{deluxetable}{cccccc}
\label{table:1}
\tablecolumns{6}
\tablewidth{0pc}

\tablecaption{Assumed parameters during Photometric solution}
\tablehead{

\colhead{Parameters} & && &&\colhead{Values}   }
\startdata

g$_{1}$\,=\,g$_{2}$              &  && & & 0.32                    \\
A$_{1}$\,=\,A$_{2}$              &  && &  & 0.5                   \\
$x_{1bolo}=x_{2bolo}$      &  && & & 0.082                      \\
$y_{1bolo}=y_{2bolo}$      &  && & & 0.646                       \\
$x_{1B}=x_{2B}$            &  && & & 0.778                       \\
$y_{1B}=y_{2B}$            &  && & & 0.301                     \\
$x_{1V}=x_{2V}$            &  && & & 0.680                       \\
$y_{1V}=y_{2V}$            &  && & & 0.297                       \\
$x_{1R}=x_{2R}$            &  && & & 0.582                       \\
$y_{1R}=Y_{2R}$            &  && & & 0.295                     \\
T1                 &  && & & 5178\,K                  \\

\enddata
\end{deluxetable}

\begin{deluxetable}{cccccc}
\label{table:1}
\tablecolumns{6}
\tablewidth{0pc}

\tablecaption{Photometric solution of the LO Com}
\tablehead{

\colhead{Parameters} & && &&\colhead{Values}   }
\startdata
$i(^{\circ})$              &  && &             & $79.991\pm0.097$                   \\
$\Omega_{1}$=$\Omega_{2}$  &  && &             & $5.8630\pm0.0033$                    \\
$T_{2}$(K)                 &  && &             & $4874\pm3$                   \\
$q(M_{2}/M_{1}) $          &  && &             & $2.478\pm0.012$                       \\
$L_{1}/(L_{1}+L_{2})_{B}$      &  && &         & $0.4069\pm0.0052$                      \\
$L_{1}/(L_{1}+L_{2})_{V}$      &  && &         & $0.3842\pm0.0039$                    \\
$L_{1}/(L_{1}+L_{2})_{R}$      &  && &         & $0.3669\pm0.003$                      \\
$r_{1}$(pole)                  &  && &         & $0.2873\pm0.0018$              \\
$r_{1}$(side)                  &  && &         & $0.3002\pm0.0022$              \\
$r_{1}$(back)                  &  && &         & $0.3358\pm0.0037$              \\
$r_{2}$(pole)                  &  && &         & $0.4385\pm0.0016$              \\
$r_{2}$(side)                  &  && &         & $0.4694\pm0.0021$              \\
$r_{2}$(back)                  &  && &         & $0.4986\pm0.0028$              \\
$f_{over}$($\%$)                 &  && &       &  $3.2\pm0.25$                  \\
$\sum W(O-C)^{2}$                 &  && &      &$0.000259$                    \\

\enddata
\end{deluxetable}

\begin{figure}
\begin{center}
 \includegraphics[bb=40 60 770 580,height=15.cm,width=20.cm,clip,angle=0,scale=0.5,angle=0]{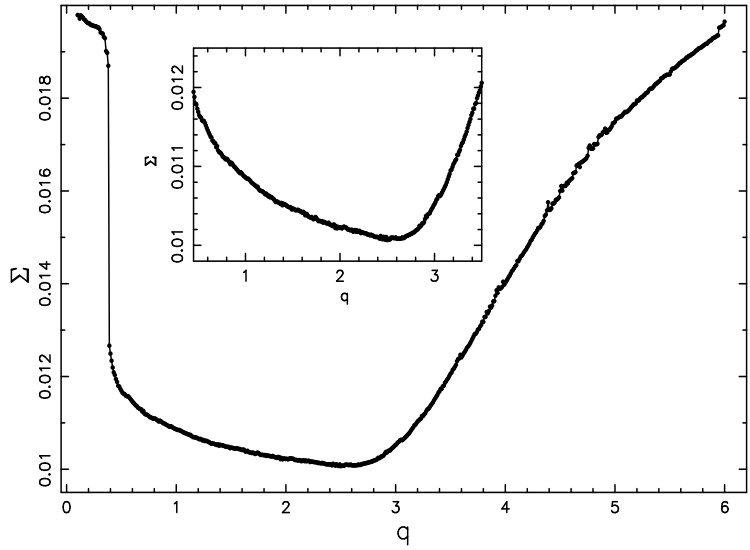}
 \caption{Relation between $\Sigma W(O-C)^{2}$ and $q$ of the LO Com.}
 \label{fig:photoz_com}
\end{center}
\end{figure}

\begin{figure}
\begin{center}
 \includegraphics[bb=38 35 710 530,height=15.cm,width=20.cm,clip,angle=0,scale=0.5,angle=0]{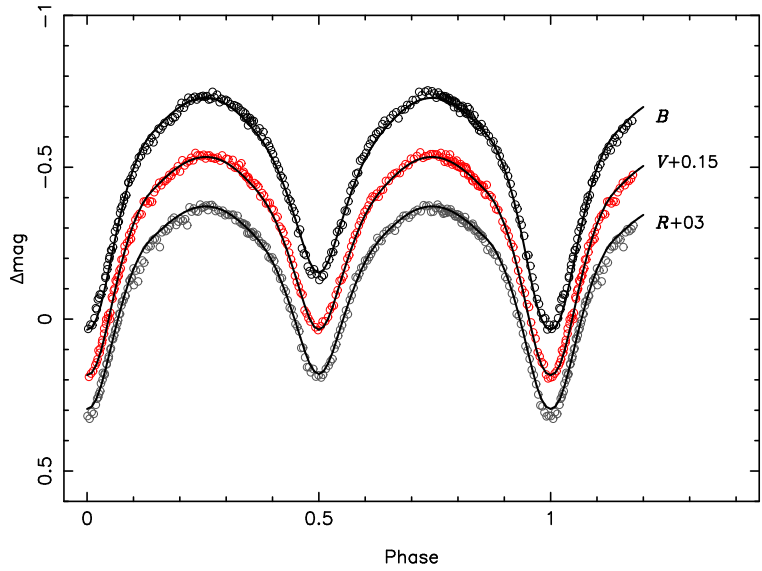}
 \caption{Observations (open circles) and computed (solid lines) light curves based on the W-D method for the $B, V$, and $R$ passbands. Note that the magnitudes and curves in the $V$ and $R$ passbands are shifted by $+0.15$ and $+0.3$ mag, respectively. }
 \label{fig:photoz_com}
\end{center}
\end{figure}
\section{Photometric solution for the LO Com }
The light curves shown in Figure 2 are symmetric and complete, all these are very useful for determining a reliable photometric solution for the LO Com. In order to derive the photometric parameters of the LO Com, the $2010$ version of the W-D program \citep{Wilson1971ApJ,Wilson1979ApJ,Wilson1990ApJ,VanHamme2007ApJ,Wilson2008ApJ} has been adopted. The spectroscopic observation can offer a reliable mass ratio. Since this system is relatively faint, minimal information about its radial velocity is available. We thus adopted thus only photometric data in our analyses. During the procedure of modeling the photometry, the effective temperature of star 1 (the star eclipsed at the primary light minimum) should be specified. Since the light curve only offers constraints on the temperature ratio by using the depth of eclipses, rather than constraints on the temperature, we estimated the temperature of star 1 based on ($J-H$) color by the expression \citep{Collier2007MNRAS}

\begin{equation}
T_{\rm eff}=-4369.5\emph{(J$-$H)}+7188.2 \ \ \ \  (4000\emph{K} < \emph{T}\rm_{eff} < 7000\emph{K}),
\end{equation}
where $J$ and $H$ magnitudes are obtained from the 2MASS All-Sky Catalog of point sources \citep{Cutri2003yCat}. From Equation (1), the effective temperature $T_1$ of the hotter component in the LO Com is derived to be of $5178$\,K. This suggests that the LO Com is a late-type W UMa system with a convective envelope, implying that the gravity-darkening coefficients $g_{1}=g_{2}=0.32$ \citep{Lucy1967} and the bolometric albedos $A_{1}=A_{2}=0.5$ \citep{rucinski1969AcA}. Note that the W-D program itself can automatically calculate the bolometric and passband limb-darkening coefficients taken from \citet{vanHamme1993AJ}, and we choose the logarithmic functions ( listed in Table 3).
The adjusted parameters are: the orbital inclination $i$, the temperature of star 2 $T_{2}$, the dimensionless surface potential $\Omega_{1}$ and $\Omega_{2}$, and the monochromatic luminosities of star 1 $L_{1B}, L_{1V}$ and $L_{1R}$.

Since so far no photometric and spectroscopic solutions have been published for the LO Com, it was necessary to obtain an accurate photometric solution for it. A series of fixed values of the mass ratio $q$ in the range of $0.1-6.0$ by step of $0.01$ were used. For each assumed mass ratio, different models (detached, semi-detached, contact, near-contact models) were tested, and the solutions usually converged to the model 3 (a contact configuration) during the computation. Therefore, a series of the sum of the weighted squared residuals, $\Sigma W (O-C)^{2}$ (hereafter $\Sigma$), were obtained. The relation between the resulting $\Sigma$ and the assumed $q$ is shown in Figure 3. It can be seen that the $\Sigma$ has a minimum value at $q\,=\,2.5$. Then, we chose $2.5$ as the initial value of $q$ and took it as an adjustable parameter in subsequent calculations until a convergent solution was obtained. Finally, $q$ was convergent to $2.478$. The parameters of the final photometric solution are listed in Table 4.

The theoretical light curves also have been obtained with the light curve program. The theoretical light curves (solid lines) together with the observations (open circles) are plotted in Figure 4. It can be seen that the theoretical light curves fit to the observed data very well. Figure 5 also shows the geometrical structure at four ($0.00, 0.25, 0.50$, and $0.75$) different orbital phases. The photometric solution shows that the LO Com has a lower contact degree of $f\,=\,3.2\%$.

For the case of poor spectroscopic data about the LO Com, we estimated the absolute parameters by the $J-H$ color. The $J-H$ color indicates that the spectral type of the LO Com is K0; its mass can be estimated to be M$_{2}=0.79$ $M_{\odot}$ \citep{Cox2000}. Then, based on the mass ratio $q\,=\,2.478$, the mass of the primary component can be estimated as approximately M$_{1}=0.32$ $M_{\odot}$.

\begin{figure}
\begin{center}
 \includegraphics[bb=90 60 700 570,height=15.cm,width=20.cm,clip,angle=0,scale=0.5,angle=0]{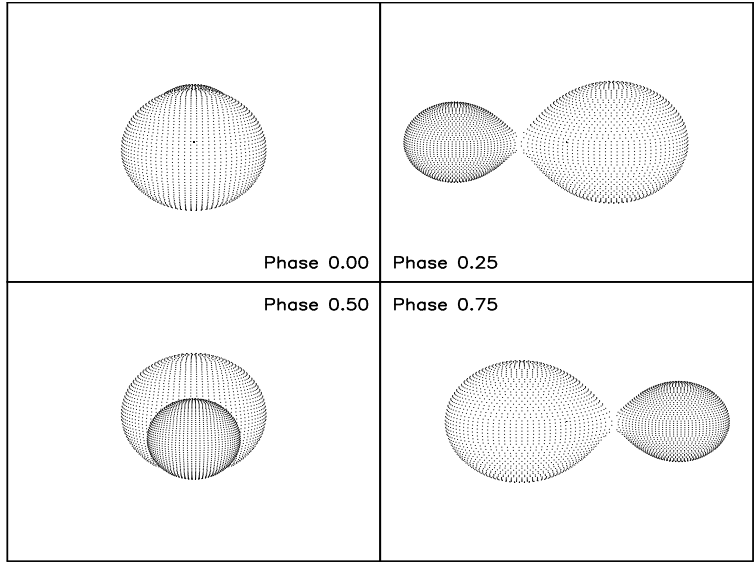}
 \caption{Configurations at the phases $0.00, 0.25, 0.50$, and $0.75$.}
 \label{fig:photoz_com}
\end{center}
\end{figure}
\begin{deluxetable}{lcccccccccccccc}
\label{table:1}
\tablecolumns{6}
\tablewidth{0pc}

\tablecaption{The CCD Times of Minimum Light of the Lo Com}
\tablehead{

\colhead{HJD+240 0000} & \colhead{Error} & \colhead{Min.}&\colhead{HJD+240 0000}& \colhead{Error} & \colhead{Min.} &\colhead{HJD+240 0000}& \colhead{Error} & \colhead{Min.}   }
\startdata
 51962.90980&    0.0001 &  s& 53813.80450&    0.0001 &  p& 54955.66630&    0.0011 &  s&\\
 51968.49456&    0.0023 &  p& 53863.34440&    0.0003 &  p& 54996.47710&    0.0007 &  s&\\
 51968.64149&    0.0023 &  s& 53863.48740&    0.0015 &  s& 55276.68070&    0.0013 &  p&\\
 51992.83460&    0.0001 &  p& 54154.57610&    0.0008 &  s& 55310.47080&    0.0025 &  p&\\
 52001.42664&    0.0022 &  p& 54170.46640&    0.0005 &  s& 55609.85640&    0.0005 &  p&\\
 52001.57082&    0.0025 &  s& 54170.61198&    0.0001 &  s& 55650.37660&    0.0003 &  s&\\
 52691.55790&    0.0017 &  p& 54170.61244&    0.0005 &  s& 55662.40470&    0.0004 &  p&\\
 53068.40780&    0.0013 &  p& 54174.47730&    0.0001 &  s& 55662.54750&    0.0007 &  p&\\
 53095.46780&    0.0015 &  s& 54174.62050&    0.0027 &  p& 55675.43360&    0.0016 &  p&\\
 53095.61140&    0.0003 &  p& 54185.35970&    0.0005 &  p& 55675.57660&    0.0006 &  s&\\
 53410.60870&    0.0005 &  p& 54200.39530&    0.0010 &  s& 55982.84060&    0.0004 &  s&\\
 53450.41200&    0.0004 &  p& 54564.36060&    0.0004 &  s& 55982.98560&    0.0007 &  s&\\
 53462.43960&    0.0017 &  p& 54594.42680&    0.0006 &  s& 56046.69820&    0.0003 &  s&\\
 53462.58200&    0.0010 &  s& 54865.89450&    0.0005 &  s& 56061.44870&    0.0020 &  s&\\
 53464.44400&    0.0030 &  p& 54933.47830&    0.0007 &  s& 56696.88000&    0.0003 &  p&\\
 53476.46950&    0.0011 &  p& 54933.62150&    0.0003 &  p& 57103.36820&    0.0002 &  s&\\
 53788.46090&    0.0011 &  s& 54937.48670&    0.0009 &  s& 57107.23430&    0.0002 &  p&\\
 
 \enddata
\end{deluxetable}
\begin{figure}
\begin{center}
 \includegraphics[bb=10 240 610 680,height=16.cm,width=20.cm,clip,angle=0,scale=0.5,angle=0]{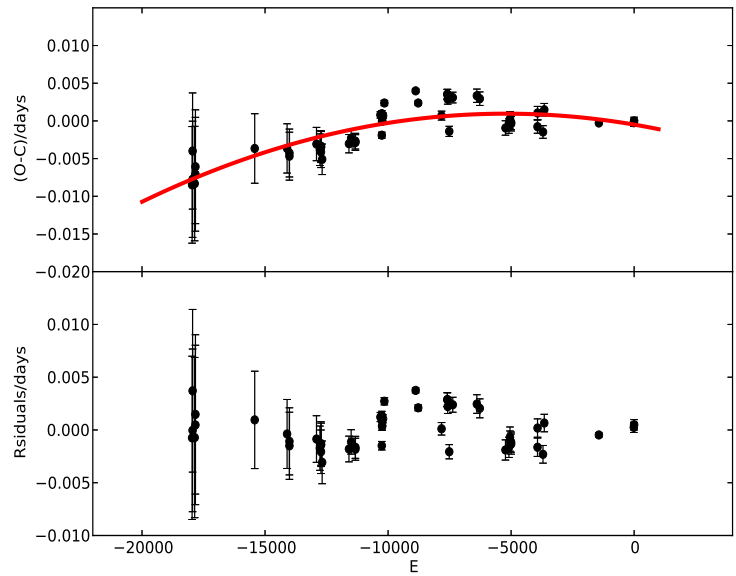}
 \caption{Upper: The $(O-C)$ diagram for all minimum times of the LO Com. The dots are obtained by the linear ephemeria in Equation (2), whereas the red solid line indicates a long-term decrease. Here the red solid line indicates a cyclic variation. The residuals after the substraction of  the best fit (the red solid line in the upper panel) are displayed in the bottom panel. }
 \label{fig:photoz_com}
\end{center}
\end{figure}

\section{Orbital period variation for the LO Com }
The orbital period change is one of the observational properties of binary stars. The study of the orbital period change for the LO Com is absent in the literature. For investigating this property, we collect the light minimum times from eclipsing binary minima database (O-C) Gateway.\footnote{http://var.astro.cz/ocgate/} Among these minimum times, only two are visual observations, and we reject these visual observations and select only the photoelectric or CCD observations. Including our new minimum times, a total number of $49$ minimum times were obtained, and the period spans over $15$ years (listed in Table 5).
Based on one of our primary times of light minimum (2457107.2343
) and the period 0.2863601 given by \citet{Blattler2001IBVS}, a new corrected linear ephemeris was obtained:
\begin{equation}
Min.I = 2457107.23766(4)+0.28636058(3) \times E
\end{equation}
The $(O-C)$ values are obtained based on this new linear ephemeris, and the relevant $(O-C)$ diagram is shown in the upper panel of Figure 6. A long-term period decrease can be seen clearly. At first, the $(O-C)$ data were fitted by a parabola; the fitted result is
\begin{equation}
%(O-C)_{1}=c_{0}+c_{1}E+c_{2}E^{2},
Min.I = 2457107.23717(\pm0.00088) + 0.28636003(\pm0.00000019)\times E-5.342(\pm0.5)\times 10^{-11}\times E^{2}.
\end{equation}
In the top panel of Figure 6, the red dashed line shows the fitted parabola. Equation $(3)$ implies that the orbital period of the LO Com decreases secularly at a rate of ${\rm d}P/{\rm d}t=-1.18\times 10^{-7}$days yr$^{-1}$.

\section{Discussions and conclusions }
In this paper, the first complete photometric light curves of the $B, V$ and $R$ passbands were presented. The photometric solution indicates that the LO Com is a W-type W UMa binary system with a contact degree of $f\,=\,(3.2 \pm 0.25)\%$. The orbital period variations were also studied, and a secular decrease in its orbital period has been found. The orbital period decreases with a rate of ${\rm d}P/{\rm d}t = -1.18\times10^{-7}$ days yr$^{-1}$.

The long-term decrease of the orbital period of the LO Com may be caused by angular momentum loss (AML) due to the magnetic stellar wind. According to the formula given by \citet{Guinan1988}:

\begin{equation}
(\frac{dP}{dt})_{\rm {AML}} \approx  -1.1\times 10^{-8}q^{'-1}(1+q^{'})^{2}(M_{1}+M_{2})^{-\frac{5}{3}}k^{2}\times(M_{1}R_{1}^{4}+M_{2}R_{2}^{4})P^{-\frac{7}{3}},
\end{equation}
the decreasing rate of the period can be estimated as ${\rm d}P/{\rm d}t\rm_{(AML)}=-4.92\times10^{-8}$days yr$^{-1}$. Here, the $M_{1}, M_{2}, R_{1},$ and $R_{2}$ in Equation (4) are the masses and radii of two components in solar units, respectively. Note that the mass ratio $q^{'}=1/q=M_{1}/M_{2}<1.0$; the $k^{2}$ was set to be $0.1$. 

The result given above seems to imply that the AML can explain the long-term period decrease for the LO Com.
However, the long-term period decrease of the LO Com might also be caused by the mass transfer from the more massive component to the less massive one by the well-known equation
\begin{equation}
\frac{\dot{P}}{P}= 3\dot{M}_{2}(\frac{1}{M_{1}}-\frac{1}{M_{2}}).
\end{equation}
Therefore, the mass transfer rate can be estimated as ${\rm d}M_{2}/{\rm d}t=-0.9\times10^{-7}$M$_{\odot}$ yr$^{-1}$. The minus sign indicates that the more massive component loses its mass. Then, the time scale of the mass transfer can be estimated to be as approximately $8.7\times10^{6}$yr. The thermal time scale of the massive component can be estimated as $\tau_{th}\sim GM_{2}/R_{2}L_{2}\sim 2.8\times10^{7}$yr \citep{Paczynski1971}, which is longer than the mass transfer duration. This implies that the primary component cannot stay in the thermal equilibrium and the mass transfer in the LO Com is unstable.

However, a comprehensive understanding of orbital period variation for the LO Com comes from the coupled contributions of the results of mass transfer from the more massive component to the less massive one, and the AML is due to magnetic stellar winds. Although, the contribution of AML is a minor aspect ($[-4.92\times10^{-8}/-1.18\times10^{-7}]< 50\%$) for the interpretation of this long-term decrease tendency. We need to note that the presence of the AML in LO Com may play an important role for the evolutionary stages from initially detached binary into overcontact binary systems \citep[e.g.,][]{Mestel68,vant79,Vilhu82,Eggen89} for the shallow contact degree ($f\,=\,(3.2 \pm 0.25)$) shown in this paper. Meanwhile, we can also see that the time scale of the secular decrease trend is about $T_{P} \sim P/({\rm d}P/{\rm d}t) \sim 2.4\times10^{6}$yr, which is close to the time scale of the mass transfer $\sim 8.7\times10^{6}$yr. This indicates that the long-term decrease in the orbital period of LO Com may be caused by the thermally conservative mass transfer from the more massive component to the less massive one.

From the photometric solution, it has been found that the LO Com is a shallow contact binary system with a contact degree of $f\,=\,3.2\%$. The temperature of the primary is hotter by about $304\,$K than the secondary component. This indicates that this binary system could be in a very important evolution phase of the TRO \citep{Flannery1976,Eggleton1996,Csizmadia2004,Li2004,Li2005,Li2008}. The binary system LO Com might be at the beginning of the duration that the full energy transfer is lost, which is indicated by the observed properties (large temperature difference and unstable mass transfer). When the effective energy transfer is lost, the secondary contracts rapidly, and the mass transfers from the primary component to the secondary component. Therefore, the secondary might have a higher temperature than the primary due to the gravitational energy transformed to thermal energy for the secondary, and rapid expansion could lead to the expense of the primary. The mass transfer rate might reach the highest value in a cycle and could induce the dynamically instability of the mass transfer \citep{Li2004}.

In summary, the LO Com is a short-period shallow contact binary system. While the complete photometric light curves are presented, long-time monitoring observations (especially spectroscopic observations) are still necessary to offer more information about this system.
\acknowledgments
We thank the anonymous referee for very valuable comments that helped us to improve
the manuscript. This work is partially supported by the program of the Light in China' Western Region (LCWR, grant No. 2015-XBQN-A-02), the National Natural Science Foundation of China (grant No. 11303080), and Youth Innovation Promotion Association CAS. The CCD photometric data of LO Com were observed with the Nanshan $1.0$ m telescope of Xinjiang Astronomical Observatory.

\bibliography{zhy}

\appendix

\section[]{The observation data of LO Com}
\begin{deluxetable}{cccccccc}
\label{table:1}
\tablecolumns{4}
\tablewidth{0pc}
\tablecaption{the original photometric Data of LO Com in the $B$, $V$, and $R$ passbands.}
\tablehead{

\colhead{HJD} & \colhead{$\Delta$m($B$)}& \colhead{HJD} & \colhead{$\Delta$m($B$)}
&\colhead{HJD} & \colhead{$\Delta$m($B$)}&\colhead{HJD} & \colhead{$\Delta$m($B$)}\\
\colhead{2457,100+}&&\colhead{2457,100+}&&\colhead{2457,100+}&&\colhead{2457,100+}&
}
\startdata
$ 3.2307$&    $-0.036$&    $ 3.3832$&    $-0.310$&    $ 3.4687$&    $-0.618$&    $ 7.2534$&    $-0.319$\\
$ 3.2318$&    $-0.065$&    $ 3.3844$&    $-0.357$&    $ 7.1678$&    $-0.731$&    $ 7.2547$&    $-0.349$\\
$ 3.2331$&    $-0.081$&    $ 3.3856$&    $-0.355$&    $ 7.1691$&    $-0.728$&    $ 7.2560$&    $-0.373$\\
$ 3.2366$&    $-0.141$&    $ 3.3868$&    $-0.388$&    $ 7.1704$&    $-0.726$&    $ 7.2572$&    $-0.402$\\
$ 3.2377$&    $-0.164$&    $ 3.3880$&    $-0.422$&    $ 7.1717$&    $-0.728$&    $ 7.2585$&    $-0.408$\\
$ 3.2389$&    $-0.198$&    $ 3.3892$&    $-0.441$&    $ 7.1731$&    $-0.714$&    $ 7.2598$&    $-0.450$\\
$ 3.2401$&    $-0.227$&    $ 3.3904$&    $-0.448$&    $ 7.1745$&    $-0.719$&    $ 7.2611$&    $-0.462$\\
$ 3.2413$&    $-0.237$&    $ 3.3915$&    $-0.479$&    $ 7.1759$&    $-0.714$&    $ 7.2624$&    $-0.475$\\
$ 3.2425$&    $-0.289$&    $ 3.3927$&    $-0.482$&    $ 7.1772$&    $-0.704$&    $ 7.2637$&    $-0.498$\\
$ 3.2437$&    $-0.302$&    $ 3.3939$&    $-0.522$&    $ 7.1784$&    $-0.700$&    $ 7.2650$&    $-0.510$\\
$ 3.2448$&    $-0.321$&    $ 3.3950$&    $-0.527$&    $ 7.1799$&    $-0.702$&    $ 7.2663$&    $-0.532$\\
$ 3.2460$&    $-0.364$&    $ 3.3962$&    $-0.535$&    $ 7.1816$&    $-0.683$&    $ 7.2675$&    $-0.538$\\
$ 3.2480$&    $-0.390$&    $ 3.3976$&    $-0.547$&    $ 7.1830$&    $-0.675$&    $ 7.2688$&    $-0.554$\\
$ 3.2508$&    $-0.449$&    $ 3.3988$&    $-0.566$&    $ 7.1842$&    $-0.675$&    $ 7.2701$&    $-0.568$\\
$ 3.3223$&    $-0.664$&    $ 3.3999$&    $-0.577$&    $ 7.1859$&    $-0.657$&    $ 7.2714$&    $-0.556$\\
$ 3.3235$&    $-0.661$&    $ 3.4011$&    $-0.592$&    $ 7.1872$&    $-0.671$&    $ 7.2727$&    $-0.583$\\
$ 3.3247$&    $-0.638$&    $ 3.4024$&    $-0.595$&    $ 7.1890$&    $-0.640$&    $ 7.2740$&    $-0.597$\\
$ 3.3259$&    $-0.662$&    $ 3.4038$&    $-0.610$&    $ 7.1903$&    $-0.633$&    $ 7.2753$&    $-0.582$\\
$ 3.3271$&    $-0.654$&    $ 3.4050$&    $-0.616$&    $ 7.1916$&    $-0.642$&    $ 7.2766$&    $-0.608$\\
$ 3.3282$&    $-0.652$&    $ 3.4063$&    $-0.621$&    $ 7.1931$&    $-0.624$&    $ 7.2785$&    $-0.627$\\
$ 3.3294$&    $-0.625$&    $ 3.4077$&    $-0.642$&    $ 7.1944$&    $-0.613$&    $ 7.2797$&    $-0.624$\\
$ 3.3306$&    $-0.617$&    $ 3.4090$&    $-0.653$&    $ 7.1961$&    $-0.603$&    $ 7.2810$&    $-0.635$\\
$ 3.3318$&    $-0.628$&    $ 3.4108$&    $-0.640$&    $ 7.1974$&    $-0.582$&    $ 7.2822$&    $-0.646$\\
$ 3.3330$&    $-0.600$&    $ 3.4125$&    $-0.661$&    $ 7.1989$&    $-0.585$&    $ 7.2847$&    $-0.653$\\
$ 3.3342$&    $-0.616$&    $ 3.4143$&    $-0.673$&    $ 7.2002$&    $-0.559$&    $ 7.2859$&    $-0.650$\\
$ 3.3353$&    $-0.581$&    $ 3.4162$&    $-0.681$&    $ 7.2015$&    $-0.555$&    $ 7.2872$&    $-0.662$\\
$ 3.3366$&    $-0.586$&    $ 3.4176$&    $-0.661$&    $ 7.2028$&    $-0.537$&    $ 7.2886$&    $-0.665$\\
$ 3.3377$&    $-0.570$&    $ 3.4196$&    $-0.689$&    $ 7.2041$&    $-0.513$&    $ 7.2899$&    $-0.676$\\
$ 3.3389$&    $-0.556$&    $ 3.4216$&    $-0.691$&    $ 7.2054$&    $-0.502$&    $ 7.2913$&    $-0.694$\\
$ 3.3401$&    $-0.543$&    $ 3.4229$&    $-0.705$&    $ 7.2067$&    $-0.492$&    $ 7.2926$&    $-0.712$\\
$ 3.3413$&    $-0.522$&    $ 3.4249$&    $-0.705$&    $ 7.2080$&    $-0.471$&    $ 7.2941$&    $-0.696$\\
$ 3.3425$&    $-0.508$&    $ 3.4264$&    $-0.721$&    $ 7.2095$&    $-0.445$&    $ 7.2955$&    $-0.711$\\
$ 3.3437$&    $-0.493$&    $ 3.4283$&    $-0.722$&    $ 7.2108$&    $-0.412$&    $ 7.2968$&    $-0.729$\\
$ 3.3449$&    $-0.472$&    $ 3.4303$&    $-0.748$&    $ 7.2125$&    $-0.377$&    $ 7.2982$&    $-0.719$\\
$ 3.3460$&    $-0.465$&    $ 3.4320$&    $-0.737$&    $ 7.2138$&    $-0.363$&    $ 7.2995$&    $-0.727$\\
\hline
\colhead{HJD} & \colhead{$\Delta$m($V$)}& \colhead{HJD} & \colhead{$\Delta$m($V$)}
&\colhead{HJD} & \colhead{$\Delta$m($V$)}&\colhead{HJD} & \colhead{$\Delta$m($V$)}\\
\colhead{2457,100+}&&\colhead{2457,100+}&&\colhead{2457,100+}&&\colhead{2457,100+}&
\\
\hline
\tablehead{
\colhead{HJD} & \colhead{$\Delta$m($V$)}& \colhead{HJD} & \colhead{$\Delta$m($V$)}
&\colhead{HJD} & \colhead{$\Delta$m($V$)}&\colhead{HJD} & \colhead{$\Delta$m($V$)}\\
\colhead{2457,100+}&&\colhead{2457,100+}&&\colhead{2457,100+}&&\colhead{2457,100+}&
}

$ 3.2311$&    $-0.007$&    $ 3.3812$&    $-0.276$&    $ 3.4679$&    $-0.593$&    $ 7.2525$&    $-0.273$\\
$ 3.2322$&    $-0.047$&    $ 3.3824$&    $-0.279$&    $ 3.4691$&    $-0.590$&    $ 7.2538$&    $-0.293$\\
$ 3.2334$&    $-0.053$&    $ 3.3836$&    $-0.291$&    $ 3.4703$&    $-0.587$&    $ 7.2551$&    $-0.317$\\
$ 3.2346$&    $-0.081$&    $ 3.3848$&    $-0.320$&    $ 3.4714$&    $-0.559$&    $ 7.2564$&    $-0.346$\\
$ 3.2358$&    $-0.119$&    $ 3.3860$&    $-0.349$&    $ 7.1695$&    $-0.692$&    $ 7.2577$&    $-0.371$\\
$ 3.2370$&    $-0.134$&    $ 3.3872$&    $-0.361$&    $ 7.1708$&    $-0.666$&    $ 7.2589$&    $-0.401$\\
$ 3.2381$&    $-0.153$&    $ 3.3884$&    $-0.389$&    $ 7.1721$&    $-0.692$&    $ 7.2603$&    $-0.427$\\
$ 3.2393$&    $-0.150$&    $ 3.3896$&    $-0.408$&    $ 7.1736$&    $-0.659$&    $ 7.2615$&    $-0.438$\\
$ 3.2405$&    $-0.223$&    $ 3.3907$&    $-0.427$&    $ 7.1750$&    $-0.662$&    $ 7.2628$&    $-0.454$\\
$ 3.2417$&    $-0.244$&    $ 3.3919$&    $-0.435$&    $ 7.1763$&    $-0.654$&    $ 7.2641$&    $-0.477$\\
$ 3.2429$&    $-0.277$&    $ 3.3931$&    $-0.458$&    $ 7.1776$&    $-0.653$&    $ 7.2654$&    $-0.488$\\
$ 3.2441$&    $-0.288$&    $ 3.3943$&    $-0.473$&    $ 7.1789$&    $-0.648$&    $ 7.2667$&    $-0.497$\\
$ 3.2453$&    $-0.310$&    $ 3.3954$&    $-0.487$&    $ 7.1808$&    $-0.631$&    $ 7.2680$&    $-0.495$\\
$ 3.2464$&    $-0.346$&    $ 3.3966$&    $-0.503$&    $ 7.1821$&    $-0.650$&    $ 7.2693$&    $-0.537$\\
$ 3.2484$&    $-0.390$&    $ 3.3980$&    $-0.511$&    $ 7.1834$&    $-0.627$&    $ 7.2706$&    $-0.538$\\
$ 3.2512$&    $-0.409$&    $ 3.3992$&    $-0.524$&    $ 7.1847$&    $-0.623$&    $ 7.2718$&    $-0.531$\\
$ 3.2524$&    $-0.414$&    $ 3.4003$&    $-0.543$&    $ 7.1863$&    $-0.604$&    $ 7.2731$&    $-0.538$\\
$ 3.3215$&    $-0.630$&    $ 3.4015$&    $-0.554$&    $ 7.1876$&    $-0.608$&    $ 7.2745$&    $-0.536$\\
$ 3.3227$&    $-0.627$&    $ 3.4028$&    $-0.552$&    $ 7.1894$&    $-0.589$&    $ 7.2757$&    $-0.568$\\
$ 3.3239$&    $-0.618$&    $ 3.4042$&    $-0.563$&    $ 7.1907$&    $-0.586$&    $ 7.2770$&    $-0.581$\\
$ 3.3251$&    $-0.610$&    $ 3.4054$&    $-0.553$&    $ 7.1920$&    $-0.588$&    $ 7.2789$&    $-0.599$\\
$ 3.3263$&    $-0.596$&    $ 3.4067$&    $-0.575$&    $ 7.1935$&    $-0.564$&    $ 7.2801$&    $-0.591$\\
$ 3.3274$&    $-0.589$&    $ 3.4081$&    $-0.576$&    $ 7.1948$&    $-0.592$&    $ 7.2814$&    $-0.605$\\
$ 3.3286$&    $-0.598$&    $ 3.4099$&    $-0.605$&    $ 7.1965$&    $-0.558$&    $ 7.2826$&    $-0.605$\\
$ 3.3298$&    $-0.593$&    $ 3.4116$&    $-0.605$&    $ 7.1978$&    $-0.552$&    $ 7.2839$&    $-0.612$\\
$ 3.3310$&    $-0.586$&    $ 3.4133$&    $-0.621$&    $ 7.1994$&    $-0.528$&    $ 7.2851$&    $-0.625$\\
$ 3.3322$&    $-0.554$&    $ 3.4151$&    $-0.629$&    $ 7.2007$&    $-0.512$&    $ 7.2863$&    $-0.622$\\
$ 3.3334$&    $-0.565$&    $ 3.4166$&    $-0.632$&    $ 7.2019$&    $-0.509$&    $ 7.2876$&    $-0.635$\\
$ 3.3346$&    $-0.557$&    $ 3.4184$&    $-0.631$&    $ 7.2045$&    $-0.471$&    $ 7.2890$&    $-0.639$\\
$ 3.3358$&    $-0.531$&    $ 3.4203$&    $-0.649$&    $ 7.2058$&    $-0.449$&    $ 7.2903$&    $-0.653$\\
$ 3.3369$&    $-0.532$&    $ 3.4219$&    $-0.660$&    $ 7.2071$&    $-0.442$&    $ 7.2917$&    $-0.636$\\
$ 3.3381$&    $-0.533$&    $ 3.4237$&    $-0.657$&    $ 7.2084$&    $-0.423$&    $ 7.2931$&    $-0.673$\\
$ 3.3393$&    $-0.513$&    $ 3.4253$&    $-0.678$&    $ 7.2100$&    $-0.398$&    $ 7.2945$&    $-0.656$\\
$ 3.3405$&    $-0.509$&    $ 3.4272$&    $-0.677$&    $ 7.2113$&    $-0.387$&    $ 7.2959$&    $-0.666$\\
$ 3.3417$&    $-0.482$&    $ 3.4290$&    $-0.687$&    $ 7.2129$&    $-0.346$&    $ 7.2972$&    $-0.670$\\
$ 3.3429$&    $-0.487$&    $ 3.4311$&    $-0.699$&    $ 7.2142$&    $-0.309$&    $ 7.2985$&    $-0.681$\\
$ 3.3441$&    $-0.455$&    $ 3.4328$&    $-0.681$&    $ 7.2155$&    $-0.277$&    $ 7.2999$&    $-0.673$\\
$ 3.3453$&    $-0.447$&    $ 3.4345$&    $-0.686$&    $ 7.2168$&    $-0.248$&    $ 7.3012$&    $-0.697$\\
$ 3.3465$&    $-0.434$&    $ 3.4361$&    $-0.688$&    $ 7.2184$&    $-0.236$&    $ 7.3025$&    $-0.689$\\
$ 3.3477$&    $-0.406$&    $ 3.4375$&    $-0.693$&    $ 7.2196$&    $-0.199$&    $ 7.3038$&    $-0.689$\\
$ 3.3489$&    $-0.396$&    $ 3.4388$&    $-0.693$&    $ 7.2209$&    $-0.190$&    $ 7.3052$&    $-0.686$\\
$ 3.3501$&    $-0.369$&    $ 3.4403$&    $-0.692$&    $ 7.2222$&    $-0.139$&    $ 7.3063$&    $-0.691$\\
$ 3.3512$&    $-0.347$&    $ 3.4414$&    $-0.692$&    $ 7.2236$&    $-0.106$&    $ 7.3075$&    $-0.679$\\
$ 3.3524$&    $-0.328$&    $ 3.4426$&    $-0.685$&    $ 7.2250$&    $-0.086$&    $ 7.3087$&    $-0.688$\\
$ 3.3536$&    $-0.293$&    $ 3.4438$&    $-0.697$&    $ 7.2263$&    $-0.041$&    $ 7.3098$&    $-0.689$\\
$ 3.3548$&    $-0.272$&    $ 3.4450$&    $-0.682$&    $ 7.2275$&    $-0.027$&    $ 7.3110$&    $-0.681$\\
$ 3.3560$&    $-0.257$&    $ 3.4462$&    $-0.668$&    $ 7.2288$&    $-0.008$&    $ 7.3122$&    $-0.681$\\
$ 3.3572$&    $-0.237$&    $ 3.4473$&    $-0.673$&    $ 7.2301$&    $-0.001$&    $ 7.3133$&    $-0.678$\\
$ 3.3584$&    $-0.231$&    $ 3.4487$&    $-0.680$&    $ 7.2314$&    $ 0.032$&    $ 7.3150$&    $-0.682$\\
$ 3.3596$&    $-0.179$&    $ 3.4500$&    $-0.669$&    $ 7.2327$&    $ 0.045$&    $ 7.3166$&    $-0.672$\\
$ 3.3608$&    $-0.169$&    $ 3.4511$&    $-0.672$&    $ 7.2341$&    $ 0.042$&    $ 7.3177$&    $-0.689$\\
$ 3.3619$&    $-0.153$&    $ 3.4524$&    $-0.667$&    $ 7.2354$&    $ 0.040$&    $ 7.3204$&    $-0.678$\\
$ 3.3631$&    $-0.142$&    $ 3.4536$&    $-0.672$&    $ 7.2367$&    $ 0.029$&    $ 7.3216$&    $-0.676$\\
$ 3.3663$&    $-0.123$&    $ 3.4548$&    $-0.651$&    $ 7.2380$&    $ 0.021$&    $ 7.3231$&    $-0.657$\\
$ 3.3679$&    $-0.114$&    $ 3.4560$&    $-0.660$&    $ 7.2393$&    $ 0.002$&    $ 7.3242$&    $-0.674$\\
$ 3.3696$&    $-0.127$&    $ 3.4572$&    $-0.653$&    $ 7.2408$&    $-0.018$&    $ 7.3256$&    $-0.649$\\
$ 3.3707$&    $-0.142$&    $ 3.4584$&    $-0.651$&    $ 7.2421$&    $-0.039$&    $ 7.3269$&    $-0.648$\\
$ 3.3724$&    $-0.143$&    $ 3.4596$&    $-0.644$&    $ 7.2435$&    $-0.072$&    $ 7.3282$&    $-0.657$\\
$ 3.3741$&    $-0.157$&    $ 3.4608$&    $-0.623$&    $ 7.2448$&    $-0.102$&    $ 7.3296$&    $-0.663$\\
$ 3.3753$&    $-0.181$&    $ 3.4619$&    $-0.627$&    $ 7.2461$&    $-0.149$&    $ 7.3310$&    $-0.635$\\
$ 3.3765$&    $-0.192$&    $ 3.4631$&    $-0.607$&    $ 7.2473$&    $-0.164$&    $ 7.3324$&    $-0.619$\\
$ 3.3777$&    $-0.195$&    $ 3.4643$&    $-0.616$&    $ 7.2486$&    $-0.171$&    $ 7.3339$&    $-0.617$\\
$ 3.3789$&    $-0.230$&    $ 3.4655$&    $-0.607$&    $ 7.2499$&    $-0.217$&    &    \\
$ 3.3800$&    $-0.245$&    $ 3.4667$&    $-0.590$&    $ 7.2512$&    $-0.250$&    &    \\
\hline
\colhead{HJD} & \colhead{$\Delta$m($R$)}& \colhead{HJD} & \colhead{$\Delta$m($R$)}
&\colhead{HJD} & \colhead{$\Delta$m($R$)}&\colhead{HJD} & \colhead{$\Delta$m($R$)}\\
\colhead{2457,100+}&&\colhead{2457,100+}&&\colhead{2457,100+}&&\colhead{2457,100+}&
\\
\hline
\tablehead{
\colhead{HJD} & \colhead{$\Delta$m($R$)}& \colhead{HJD} & \colhead{$\Delta$m($R$)}
&\colhead{HJD} & \colhead{$\Delta$m($R$)}&\colhead{HJD} & \colhead{$\Delta$m($R$)}\\
\colhead{2457,100+}&&\colhead{2457,100+}&&\colhead{2457,100+}&&\colhead{2457,100+}&
}
$ 3.2314$&    $-0.038$&    $ 3.3804$&    $-0.245$&    $ 3.4682$&    $-0.579$&    $ 7.2542$&    $-0.318$\\
$ 3.2326$&    $-0.054$&    $ 3.3816$&    $-0.265$&    $ 3.4694$&    $-0.574$&    $ 7.2555$&    $-0.345$\\
$ 3.2338$&    $-0.088$&    $ 3.3828$&    $-0.280$&    $ 7.1673$&    $-0.655$&    $ 7.2568$&    $-0.361$\\
$ 3.2350$&    $-0.123$&    $ 3.3840$&    $-0.303$&    $ 7.1686$&    $-0.650$&    $ 7.2580$&    $-0.376$\\
$ 3.2361$&    $-0.121$&    $ 3.3851$&    $-0.329$&    $ 7.1699$&    $-0.650$&    $ 7.2593$&    $-0.401$\\
$ 3.2373$&    $-0.153$&    $ 3.3863$&    $-0.316$&    $ 7.1712$&    $-0.653$&    $ 7.2606$&    $-0.427$\\
$ 3.2385$&    $-0.170$&    $ 3.3875$&    $-0.363$&    $ 7.1725$&    $-0.661$&    $ 7.2619$&    $-0.436$\\
$ 3.2397$&    $-0.200$&    $ 3.3887$&    $-0.378$&    $ 7.1754$&    $-0.639$&    $ 7.2632$&    $-0.412$\\
$ 3.2408$&    $-0.236$&    $ 3.3899$&    $-0.403$&    $ 7.1767$&    $-0.638$&    $ 7.2645$&    $-0.460$\\
$ 3.2420$&    $-0.247$&    $ 3.3911$&    $-0.438$&    $ 7.1780$&    $-0.635$&    $ 7.2658$&    $-0.424$\\
$ 3.2432$&    $-0.269$&    $ 3.3922$&    $-0.435$&    $ 7.1793$&    $-0.617$&    $ 7.2671$&    $-0.463$\\
$ 3.2444$&    $-0.296$&    $ 3.3934$&    $-0.450$&    $ 7.1812$&    $-0.623$&    $ 7.2684$&    $-0.493$\\
$ 3.2456$&    $-0.317$&    $ 3.3946$&    $-0.469$&    $ 7.1825$&    $-0.613$&    $ 7.2696$&    $-0.486$\\
$ 3.2468$&    $-0.358$&    $ 3.3958$&    $-0.472$&    $ 7.1838$&    $-0.603$&    $ 7.2709$&    $-0.530$\\
$ 3.2487$&    $-0.377$&    $ 3.3970$&    $-0.493$&    $ 7.1850$&    $-0.610$&    $ 7.2722$&    $-0.539$\\
$ 3.2515$&    $-0.427$&    $ 3.3983$&    $-0.493$&    $ 7.1867$&    $-0.606$&    $ 7.2735$&    $-0.540$\\
$ 3.2527$&    $-0.437$&    $ 3.3995$&    $-0.504$&    $ 7.1880$&    $-0.594$&    $ 7.2748$&    $-0.560$\\
$ 3.2539$&    $-0.467$&    $ 3.4007$&    $-0.515$&    $ 7.1898$&    $-0.575$&    $ 7.2761$&    $-0.565$\\
$ 3.3218$&    $-0.615$&    $ 3.4019$&    $-0.528$&    $ 7.1911$&    $-0.578$&    $ 7.2774$&    $-0.562$\\
$ 3.3230$&    $-0.609$&    $ 3.4032$&    $-0.535$&    $ 7.1924$&    $-0.565$&    $ 7.2792$&    $-0.537$\\
$ 3.3242$&    $-0.583$&    $ 3.4045$&    $-0.567$&    $ 7.1939$&    $-0.554$&    $ 7.2805$&    $-0.586$\\
$ 3.3254$&    $-0.581$&    $ 3.4057$&    $-0.559$&    $ 7.1952$&    $-0.528$&    $ 7.2817$&    $-0.596$\\
$ 3.3266$&    $-0.583$&    $ 3.4071$&    $-0.557$&    $ 7.1969$&    $-0.536$&    $ 7.2829$&    $-0.590$\\
$ 3.3278$&    $-0.587$&    $ 3.4084$&    $-0.580$&    $ 7.1982$&    $-0.521$&    $ 7.2843$&    $-0.596$\\
$ 3.3290$&    $-0.575$&    $ 3.4102$&    $-0.590$&    $ 7.1997$&    $-0.512$&    $ 7.2854$&    $-0.608$\\
$ 3.3302$&    $-0.578$&    $ 3.4119$&    $-0.595$&    $ 7.2010$&    $-0.510$&    $ 7.2866$&    $-0.627$\\
$ 3.3314$&    $-0.559$&    $ 3.4136$&    $-0.602$&    $ 7.2023$&    $-0.490$&    $ 7.2880$&    $-0.634$\\
$ 3.3326$&    $-0.550$&    $ 3.4154$&    $-0.591$&    $ 7.2036$&    $-7.584$&    $ 7.2893$&    $-0.610$\\
$ 3.3337$&    $-0.548$&    $ 3.4169$&    $-0.614$&    $ 7.2049$&    $-0.468$&    $ 7.2907$&    $-0.621$\\
$ 3.3349$&    $-0.540$&    $ 3.4188$&    $-0.628$&    $ 7.2062$&    $-0.439$&    $ 7.2920$&    $-0.629$\\
$ 3.3361$&    $-0.537$&    $ 3.4207$&    $-0.640$&    $ 7.2075$&    $-0.431$&    $ 7.2935$&    $-0.624$\\
$ 3.3373$&    $-0.505$&    $ 3.4223$&    $-0.641$&    $ 7.2088$&    $-0.409$&    $ 7.2949$&    $-0.629$\\
$ 3.3385$&    $-0.500$&    $ 3.4241$&    $-0.634$&    $ 7.2103$&    $-0.384$&    $ 7.2962$&    $-0.609$\\
$ 3.3397$&    $-0.503$&    $ 3.4256$&    $-0.644$&    $ 7.2116$&    $-0.361$&    $ 7.2975$&    $-0.668$\\
$ 3.3409$&    $-0.491$&    $ 3.4275$&    $-0.654$&    $ 7.2133$&    $-0.317$&    $ 7.2989$&    $-0.654$\\
$ 3.3420$&    $-0.464$&    $ 3.4293$&    $-0.660$&    $ 7.2146$&    $-0.311$&    $ 7.3002$&    $-0.664$\\
$ 3.3433$&    $-0.456$&    $ 3.4314$&    $-0.655$&    $ 7.2159$&    $-0.284$&    $ 7.3015$&    $-0.656$\\
$ 3.3445$&    $-0.433$&    $ 3.4331$&    $-0.675$&    $ 7.2172$&    $-0.253$&    $ 7.3028$&    $-0.654$\\
$ 3.3456$&    $-0.435$&    $ 3.4348$&    $-0.672$&    $ 7.2187$&    $-0.211$&    $ 7.3042$&    $-0.677$\\
$ 3.3468$&    $-0.401$&    $ 3.4364$&    $-0.674$&    $ 7.2200$&    $-0.197$&    $ 7.3055$&    $-0.667$\\
$ 3.3480$&    $-0.391$&    $ 3.4378$&    $-0.653$&    $ 7.2213$&    $-0.169$&    $ 7.3067$&    $-0.670$\\
$ 3.3492$&    $-0.362$&    $ 3.4391$&    $-0.654$&    $ 7.2226$&    $-0.133$&    $ 7.3078$&    $-0.674$\\
$ 3.3504$&    $-0.337$&    $ 3.4406$&    $-0.666$&    $ 7.2239$&    $-0.116$&    $ 7.3090$&    $-0.659$\\
$ 3.3516$&    $-0.333$&    $ 3.4418$&    $-0.677$&    $ 7.2253$&    $-0.086$&    $ 7.3102$&    $-0.675$\\
$ 3.3528$&    $-0.310$&    $ 3.4430$&    $-0.668$&    $ 7.2266$&    $-0.052$&    $ 7.3113$&    $-0.676$\\
$ 3.3540$&    $-0.284$&    $ 3.4442$&    $-0.664$&    $ 7.2279$&    $-0.016$&    $ 7.3125$&    $-0.669$\\
$ 3.3552$&    $-0.275$&    $ 3.4453$&    $-0.647$&    $ 7.2292$&    $-0.013$&    $ 7.3137$&    $-0.656$\\
$ 3.3563$&    $-0.260$&    $ 3.4465$&    $-0.657$&    $ 7.2305$&    $ 0.007$&    $ 7.3153$&    $-0.659$\\
$ 3.3575$&    $-0.205$&    $ 3.4477$&    $-0.663$&    $ 7.2318$&    $ 0.020$&    $ 7.3169$&    $-0.651$\\
$ 3.3587$&    $-0.206$&    $ 3.4490$&    $-0.653$&    $ 7.2331$&    $ 0.018$&    $ 7.3181$&    $-0.657$\\
$ 3.3599$&    $-0.174$&    $ 3.4503$&    $-0.652$&    $ 7.2345$&    $ 0.019$&    $ 7.3196$&    $-0.642$\\
$ 3.3611$&    $-0.142$&    $ 3.4515$&    $-0.647$&    $ 7.2358$&    $ 0.028$&    $ 7.3207$&    $-0.646$\\
$ 3.3623$&    $-0.142$&    $ 3.4528$&    $-0.639$&    $ 7.2371$&    $ 0.006$&    $ 7.3219$&    $-0.643$\\
$ 3.3635$&    $-0.125$&    $ 3.4540$&    $-0.645$&    $ 7.2384$&    $ 0.012$&    $ 7.3234$&    $-0.643$\\
$ 3.3650$&    $-0.136$&    $ 3.4552$&    $-0.642$&    $ 7.2397$&    $-0.018$&    $ 7.3246$&    $-0.644$\\
$ 3.3666$&    $-0.111$&    $ 3.4564$&    $-0.629$&    $ 7.2411$&    $-0.043$&    $ 7.3259$&    $-0.626$\\
$ 3.3683$&    $-0.115$&    $ 3.4575$&    $-0.625$&    $ 7.2424$&    $-0.063$&    $ 7.3272$&    $-0.640$\\
$ 3.3699$&    $-0.108$&    $ 3.4587$&    $-0.636$&    $ 7.2439$&    $-0.101$&    $ 7.3286$&    $-0.620$\\
$ 3.3711$&    $-0.117$&    $ 3.4599$&    $-0.612$&    $ 7.2451$&    $-0.117$&    $ 7.3299$&    $-0.614$\\
$ 3.3727$&    $-0.136$&    $ 3.4611$&    $-0.607$&    $ 7.2464$&    $-0.151$&    $ 7.3314$&    $-0.611$\\
$ 3.3744$&    $-0.147$&    $ 3.4623$&    $-0.609$&    $ 7.2477$&    $-0.177$&    $ 7.3327$&    $-0.610$\\
$ 3.3756$&    $-0.161$&    $ 3.4635$&    $-0.614$&    $ 7.2490$&    $-0.204$&    $ 7.3342$&    $-0.585$\\
$ 3.3768$&    $-0.169$&    $ 3.4647$&    $-0.591$&    $ 7.2503$&    $-0.234$&    $ 7.3357$&    $-0.585$\\
$ 3.3780$&    $-0.203$&    $ 3.4658$&    $-0.593$&    $ 7.2516$&    $-0.264$&    $ 7.3371$&    $-0.571$\\
$ 3.3792$&    $-0.217$&    $ 3.4670$&    $-0.586$&    $ 7.2529$&    $-0.285$&    &    \\
\enddata
\end{deluxetable}

\clearpage

\end{document}